\documentclass[preprint,proceedings]{rmaa}

\suppressfulladdresses 

\usepackage{paralist}

\usepackage{psfrag,color}

\SetYear{2005}
\SetConfTitle{11th Latin American IAU Regional Meeting}

\title{High Resolution OH Maser Survey in Star Forming Regions}

\author{
  A. E. Ruiz-Velasco,\altaffilmark{1} 
  V. Migenes,\altaffilmark{1}
  V. Slysh\altaffilmark{2} 
  and I. E. Val'tts\altaffilmark{2}}

\altaffiltext{1}{Departmento de Astronom\'ia, Universidad de Guanajuato.
  Apdo. Postal 144, C.P. 36000, 
  Guanajuato, Gto., M\'exico (alma@astro.ugto.mx, vmigenes@astro.ugto.mx).}

\altaffiltext{2}{Astro Space Center of the Lebedev Physical Center of RAS,
  Profsoyuznaya Str. 84/32, 117997 Moscow, Russia (vslysh@asc.rssi.ru, ivaltts@asc.rssi.ru).}

\shortauthor{Ruiz-Velasco et al.}

\shorttitle{RevMexAA(SC) OH Maser Survey}

\listofauthors{A. E. Ruiz-Velasco, V. Migenes, V. Slysh, \& L. Val'tts}

\indexauthor{Ruiz-Velasco, A. E.}
\indexauthor{Migenes, V.}
\indexauthor{Slysh, V.}
\indexauthor{Val'tts, I. E.}

\abstract{We present results of a high resolution survey of OH 
masers in Galactic Star Forming Regions  in order 
to study the maser emission and establish a list of 
suitable candidates for higher resolution instruments follow up. 
We used the Very Long Baseline Array 
(VLBA) to observe the 1665, 1667, 1612 and 1720 MHz OH 
maser transitions within 41 regions. These are the first 
high resolution observations for most of the sources. 
For all the transitions 30 sites of maser emission were 
detected, 4 of the sources have new detections,  and 
approximately 40\%  of the sources in the sample exhibit 
highly compact structure. Finally we consider that the spectrum observed in 
 W75N shows the early stage of a long period OH 
maser flare in the 1665 MHz line, the first of its 
kind.
}

\resumen{Presentamos los resultados de un rastreo de alta resoluci\'on de
 m\'aseres de OH en Regiones de Formaci\'on Estelar   Gal\'actica con el prop\'osito
 de estudiar la emisi\'on m\'aser y establecer una lista de candidatos adecuados 
 para realizar un seguimiento con instrumentos de mayor resoluci\'on.
  Se utiliz\'o  el \textit{Very Long Baseline Array} (VLBA) para
 observar las transiciones del m\'aser de OH en 1665, 1667, 1612 y 1720 MHz dentro de 
 41 regiones. \'Estas son las primeras observaciones de alta resoluci\'on 
 que se realizan en la mayor parte de las fuentes.
 Se detectaron 30 sitios de emisi\'on m\'aser en alguna o varias transiciones,
 con nuevas detecciones en 4 fuentes,   
 y  donde aproximadamente 40\% de la muestra exhibe  estructura muy compacta.
 Finalmente consideramos que el espectro observado en  W75N muestra
 el estado inicial de una r\'afaga del m\'aser de OH en 
 la l\'inea de 1665 MHz, la cual es la primera que se conoce}

\addkeyword{masers}
\addkeyword{polarization}
\addkeyword{Stars: formation}

\begin{document}

\maketitle

\section{Introduction}
\label{sec:intro}

In this paper we present results of a VLBA survey of OH masers in
galactic Star Forming Regions (SRF) to study the high angular and spectral
resolution characteristics
of the maser emission, and to establish a list
of suitable candidates for higher resolution instruments. Due to the
limited sensitivity of the large VLBI baselines only very strong and
compact sources are detected.

\section{Observations}
\label{sec:obs}
We selected 41 galactic radio sources, most of them previously known to be high mass SFR or compact HII
regions. Observations were conducted in January 2001 with
the NRAO's
\footnote{The National Radio Astronomy Observatory is a facility of the National
Science Foundation operated under cooperative agreement by Associated
Universities, Inc.
} Very Long Baseline Array using all the 10 antennas.

The data was observed with a 250 kHz band width and processed with
256 channels so a spectral resolution of 0.98 KHz $(0.17\, km\, s^{-1})$
was achieved. We observed four OH maser transitions: 1665, 1667, 1612 and 1720 MHz,
in RCP and LCP polarization modes. We used the VLBA
in a snap-shot mode, observing every source with a 6 minutes scan,
providing a sensitivity of $\sim60\, mJy\, beam^{-1}$ and an angular
resolution of 4.3 mas. The data were processed using the NRAO's AIPS
software package with the spectral line VLBI data reduction standard
protocol.

\begin{figure*}[!t]

\includegraphics[scale=0.9] {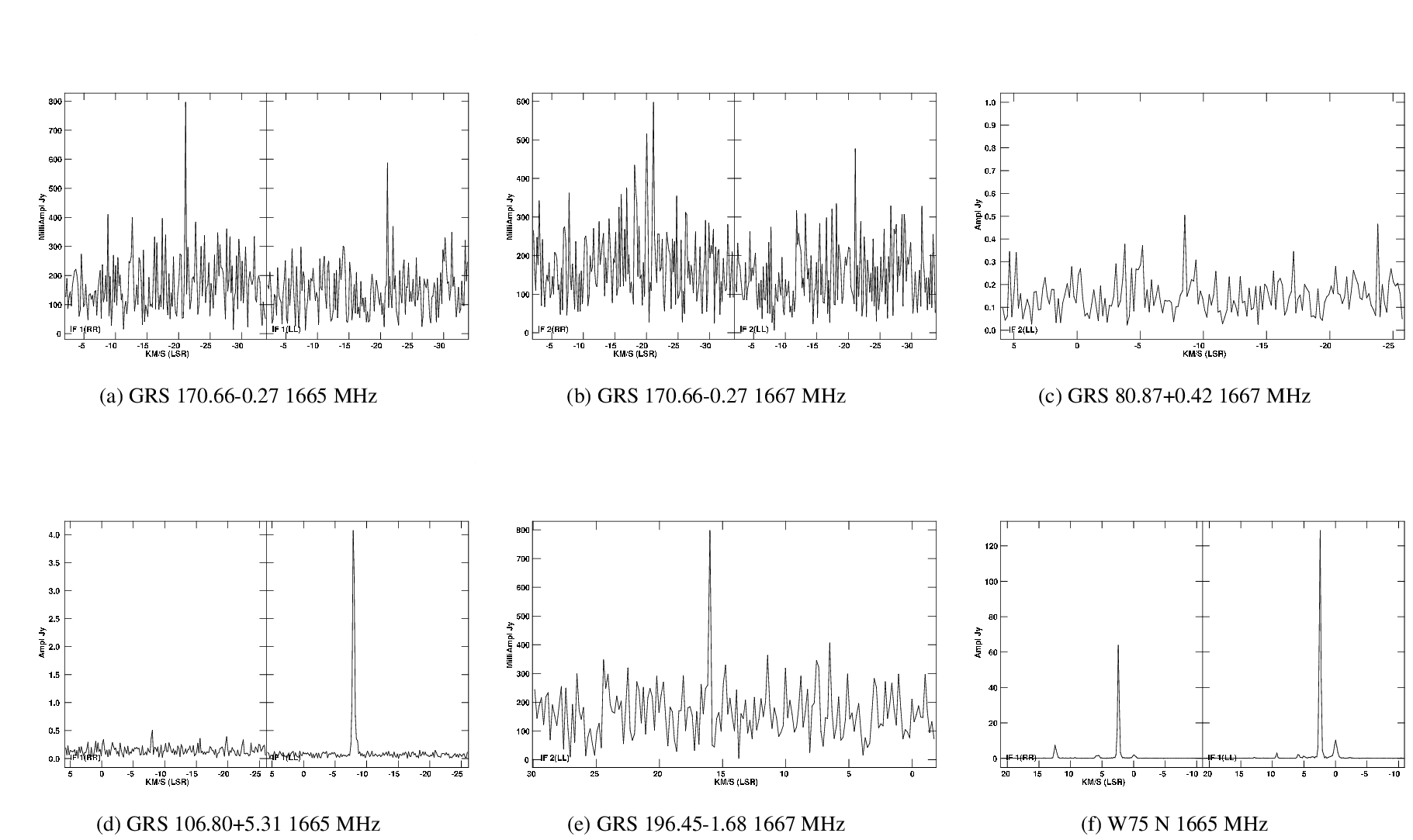}
 
\caption{Spectra of the sources with new detections (a), (b), (c), (d), (e) and the 
progenitor of the flare in W75 N (f).}

\end{figure*}

\section{Results}
\label{sec:res}

We obtain spectra for a total of 30  sources, fifteen with high compact
structure and strong emission.

We measured the corresponding magnetic
field for 3 Zeeman pairs: $-$2.86 mG in GRS 10.62-0.38 (1667 MHz),
3.93 mG in GRS 12.89+0.49 (1665 MHz) and 2.4 mG in Cepheus A (1667
MHz).

The spectra of some sources are shown in Fig.1. The spectra
are composed of an average over all baselines within the specified
range. The name and frequency observed for each source
is indicated in a sub-caption located under the spectrum panel. The
horizontal axis shows the LSR radial velocity in $km\, s^{-1}$ and the vertical
axis shows the flux intensity in Janskys. 

We found new detections in four sources: \emph{GRS 170.66-0.27} (Fig. 1a and 1b), 
\emph{GRS 80.87+0.42} (Fig. 1c), \emph{GRS 106.80+5.31} (Fig. 1d) and 
\emph{GRS 196.45-1.68} (Fig. 1e).

\subsection{OH maser flare (W75N)}

W75 is a well known region of active high-mass star formation located at a distance $\leq2\, kpc$
(Odenwald \& Schwartz 1993). It contains three main radio continuum
sources, one of them (VLA 2) harboring a UCHII region (Torrelles et
al. 1997)

The 1665
MHz spectrum show 5 strong features spread over a velocity range of
15 $km\, s^{-1}$, where the stronger has a flux of
130 Jy (LCP) and 62 Jy (RCP) at the radial velocity of 2.4 $km\, s^{-1}$
(Fig. 1f).

A flare of radio emission in the 1665 MHz line was discovered in 2003, 
with a flux density of $> 1000$ Jy (Alakoz et al, 2005). 
Its radial velocity of $~2\,km\, s^{-1}$ and position
means it is associated with VLA 2. This is the first OH maser flare ever reported,
 with an increase in the
intensity by a factor of 7 in a period of $< 3$ years. 
The emission has a high degree of linear polarization $( > 80\%)$. 

In our VLBA spectrum we identify the precursor of the flare, that
 appears to come from the same VLA 2 radio source.

\section{Conclusions and Future Work}
\label{sec:conc}

A high resolution OH maser survey was made with the VLBA over 41 galactic
SFR. We obtained high resolution spectra for 30 of them.

We report the discovery of five new maser emission sites: two in the
1665 MHz line, three in the 1667 MHz. 

We observed a weak flare in the source W75 N, proposed to be a precursor
of the powerful flare occurred during the summer of 2004: The brightest
OH maser flare ever reported. We are preparing new interferometric
observations for this source with the EVN.

\acknowledgements{V. M. and A. R. V acknowledges the support of CONACyT, M\'exico. This
research has made use of the SIMBAD database, operated at CDS, Strasbourg,
France.}


\begin{thebibliography}

\bibitem{} Alakoz, A. V., Slysh, V. I., Popov, M. V., \& Val' tts I. E. 2005,
Astronomy Letters in press, astro-ph/0501539

\bibitem{} Odenwald, S. F., \& Schwartz, P. R. 1993, ApJ, 405, 706 

\bibitem{} Torrelles, J. M., G\'omez, J. F., Rodr\'iguez, L. F., Ho, P. T., Curiel,
S., \& V\'azquez, R. 1997, ApJ, 489, 744 


\end{thebibliography}
\end{document}